\documentclass[aps,superscriptaddress,amsmath,amssymb,showpacs,twocolumn]{revtex4}

\usepackage{graphicx}
\usepackage{epstopdf}
\usepackage{hyperref}
\usepackage{epsfig}
\usepackage{color}
\usepackage{psfrag}
\usepackage{float}
\usepackage[squaren,Gray]{SIunits}
\usepackage{bbold}
\usepackage{tikz}
\usepackage[caption=false]{subfig}

\begin{document}

\title{Classical decoherence in a nanomechanical resonator}

\author{O. Maillet}

\affiliation{
Institut N\'eel, CNRS and Universit\'e Grenoble Alpes, \\
BP 166, 38042 Grenoble Cedex 9, France}

\author{F. Vavrek}

\affiliation{
Centre of Low Temperature Physics, Institute of Experimental Physics, S.A.S. and P.J. \v{S}af\'arik University Ko\v{s}ice, Watsonova 47, 04001 Ko\v{s}ice, Slovakia}

\author{A. D. Fefferman}
\affiliation{
Institut N\'eel, CNRS and Universit\'e Grenoble Alpes, \\
BP 166, 38042 Grenoble Cedex 9, France}

\author{O. Bourgeois}

\affiliation{
Institut N\'eel, CNRS and Universit\'e Grenoble Alpes, \\
BP 166, 38042 Grenoble Cedex 9, France}

\author{E. Collin}
\affiliation{
Institut N\'eel, CNRS and Universit\'e Grenoble Alpes, \\
BP 166, 38042 Grenoble Cedex 9, France}

\begin{abstract}
Decoherence is an essential mechanism that defines the boundary between classical and quantum behaviours, while imposing technological bounds for quantum devices. 
 Little is known about quantum coherence of mechanical systems, as opposed to electromagnetic degrees of freedom. 
But decoherence can also be thought of in a purely classical context, as the loss of phase coherence in the classical phase space. 
Indeed the bridge between quantum and classical physics is under intense investigation, using in particular classical nanomechanical analogues of quantum phenomena. 
In the present work, by separating pure dephasing from dissipation, we quantitatively model the classical decoherence of a mechanical resonator: through the experimental control of frequency fluctuations, we engineer artificial dephasing. 
Building on the fruitful analogy introduced between spins/quantum bits and nanomechanical modes, we report on the methods available to define pure dephasing in these systems, while demonstrating the intrinsic almost-ideal properties of silicon-nitride beams. These experimental and theoretical results, at the boundary between classical nanomechanics and quantum information fields, are prerequisite in the understanding of decoherence processes in mechanical devices, both classical and quantum.
\end{abstract}


\newcommand{\Teff}{D} 

\maketitle

\section{INTRODUCTION}
\vspace*{-3mm}
\label{intro}

Decoherence can be viewed either in its quantum picture, where it stands for the loss of phase coherence of a superposition state \cite{PRLArmour, PRBRemus}, or as its classical equivalent, where the phase of an oscillating signal is smeared due to frequency fluctuations \cite{JAPCleland}. Until recently, dissipation, which accounts for an energy loss over time \cite{PRLUnterreithmeier}, was not distinguished from decoherence in nanomechanical resonators. The first reason for this oversight lies in the acquisition method: spectral acquisition, in which the driving frequency is swept through the mechanical resonance, does not separate the two processes, since both lead to broadened resonance lines. The response linewidth then leads to the definition of a decoherence time $T_2$, in analogy with nuclear magnetic resonance (NMR) \cite{abragam} or quantum information experiments \cite{PRBIthier,NatCommsSchneider}. One way to unravel a dephasing mechanism is to perform a complementary time-domain ringdown measurement. Here, the free decay of the response amplitude yields an energy relaxation time, since the decay rate only depends on dissipation, not frequency fluctuations. This procedure is analogous to a $T_1$ measurement for spins or qubits. Having $T_1\neq T_2$ then leads to the definition of a  pure dephasing rate $\Gamma_{\phi}$ \cite{NatCommsSchneider}.

While the reported $T_1$ and $T_2$ for a particular qubit are often significantly different \cite{PRBIthier}, such a difference in a nanomechanical resonator is still rare in the literature: usually mechanical systems seem to experience frequency fluctuations small compared to dissipation mechanisms, and do not exhibit visible spectral broadening due to dephasing \cite{NatOConnell, NatPhysFaust,PRLFaust, APLVanLeeuwen}. 
Nonetheless, pure dephasing has been observed in Refs. \cite{PRBFong,NanoLettMiao,PRLZhang,NatNanoMoser}. But direct comparisons between time-domain ($T_1$) and frequency-domain ($T_2$) are still rare: thus, the second and main reason for the scarcity of experimental studies on mechanical decoherence is a lack of data {\it combining} these techniques.
To our knowledge, only one recent work reports, for a suspended carbon nanotube, a signature of a pure (yet nonlinear) dephasing mechanism demonstrating $T_1\neq T_2$ \cite{NatCommsSchneider}, while another study which {\it did not} contain $T_1$ measurements presents similar features interpreted as nonlinear damping \cite{NatNanoEichler}.
Indeed, the effects of frequency noise (and its origin) in nanomechanical devices is a subject intensely investigated today, both for its fundamental aspects and the technical limitations it poses to actual devices \cite{ArXivSansa,APLGray,PRBMaizelis,PRBFong,NanoLettMiao,
PRLZhang, NatCommsEichler, NatNanoMoser,NatCommsGavartin, NatPhysGieseler, Telegraph}. Besides, with the advances of quantum nanomechanics, the decoherence of quantum mechanical states also becomes a challenging issue \cite{NatOConnell,NJPKleckner,ArmourNJP,BowNJP,NatVerhagen}.

In this article we report on a model experiment using a high quality nanoelectromechanical system (NEMS), top-down fabricated from high-stress silicon-nitride (SiN) \cite{PRLUnterreithmeier,PRBFong}. 
Using a gate electrode \cite{APLKozinsky}, we capacitively  control the oscillator's frequency fluctuations by applying voltage noise, leading to a completely quantitative description of mechanical decoherence. 
First we perform time and frequency domain measurements without noise applied on the gate, in order to ensure that no pure dephasing is observed over our whole dynamic range, from linear to highly nonlinear regimes: we thus establish our device as ideal 
 for our study. In a second part, a low-frequency noise gate voltage is injected, leading to resonance frequency fluctuations 
  increasing with the noise level: indeed, a significant spectral broadening is observed, while ringdown measurements leave $T_1$ unchanged, demonstrating 
pure dephasing with $T_1\neq T_2$. We present the complete formalism applying to low-frequency fluctuations of the resonance frequency. 
Building on the new methods presented, we furthermore discuss the possibility to extract information about the fluctuations statistics  from the shape of the spectral response in actual devices suffering from dephasing.

\section{THE NANOELECTROMECHANICAL SYSTEM}
\label{systpart}

\begin{figure}
	\includegraphics[width=7.2cm]{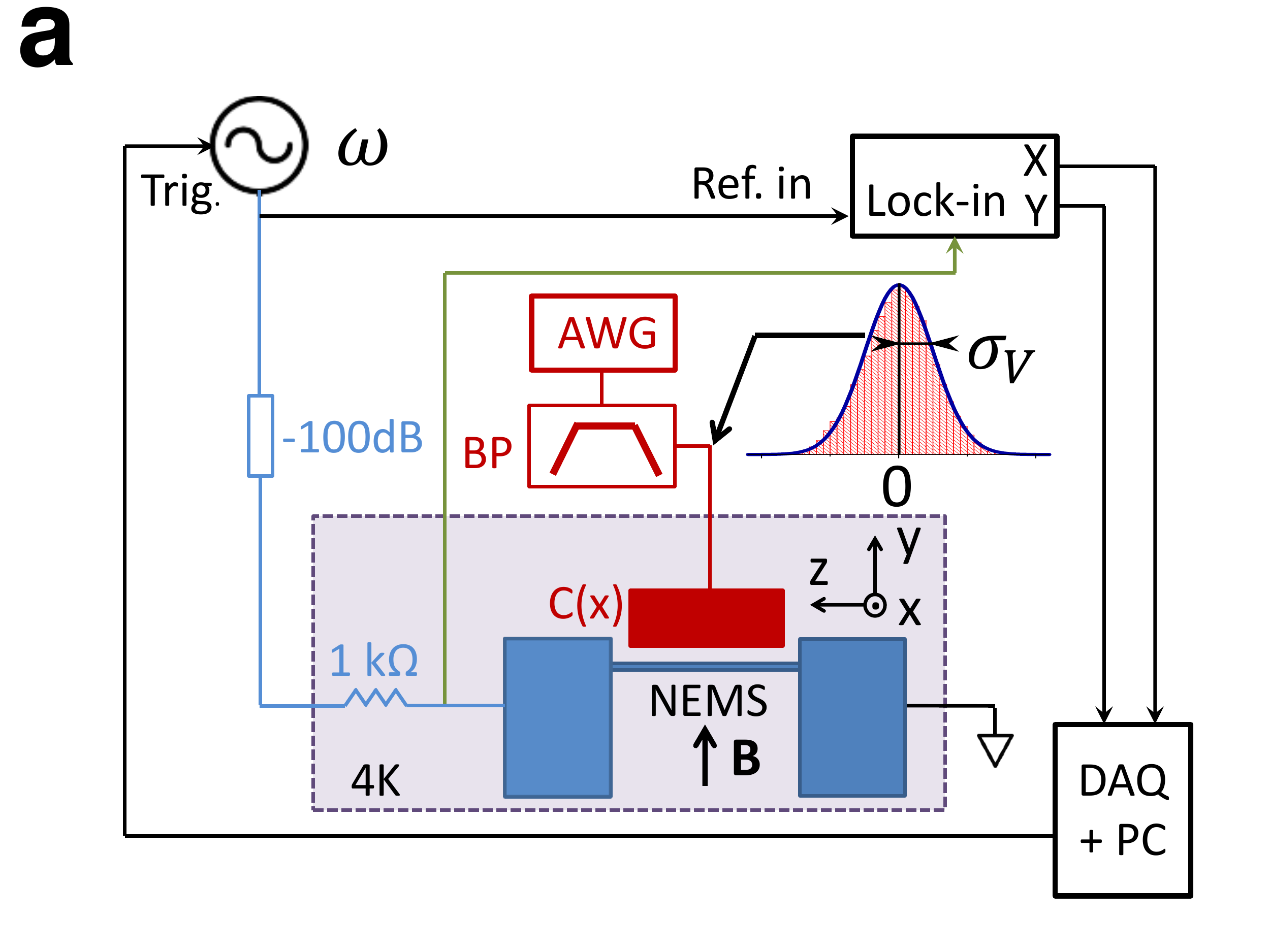}
			\includegraphics[scale=0.25]{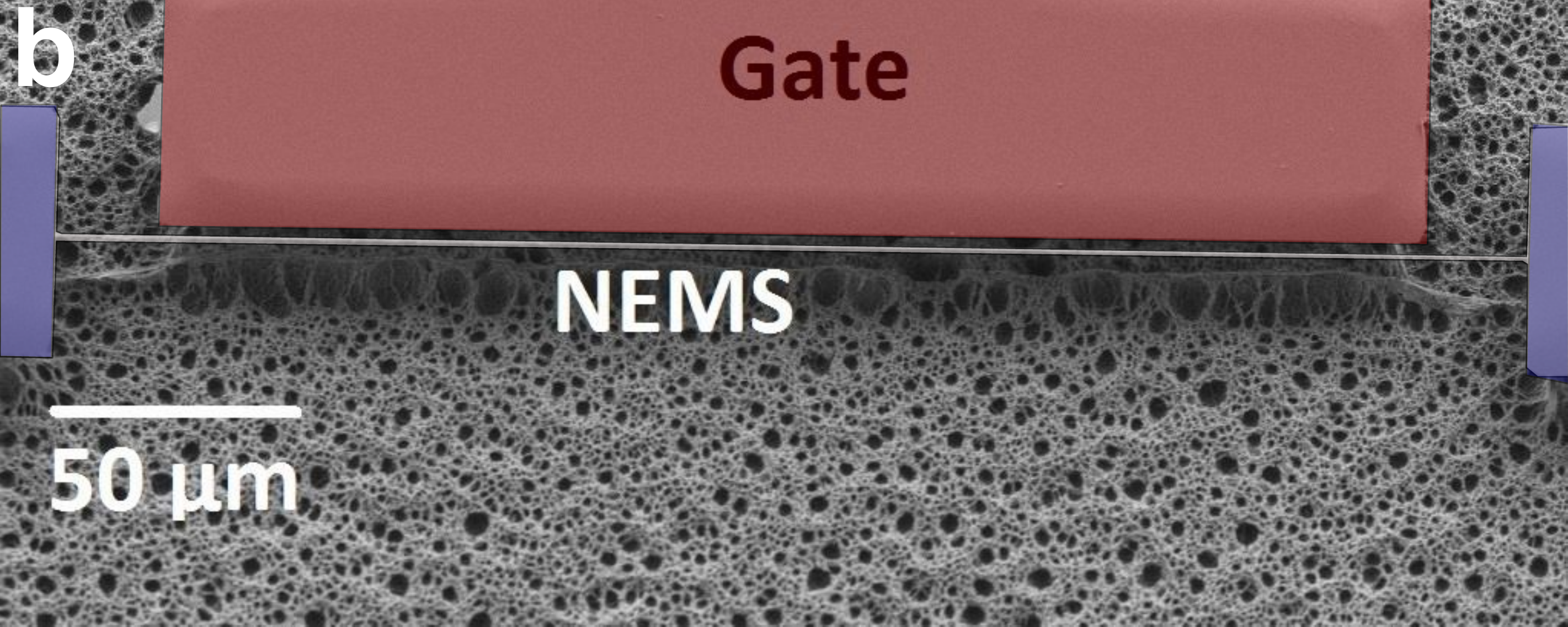}
		\begin{minipage}[c]{.45\linewidth}
		 	\includegraphics[scale=0.40]{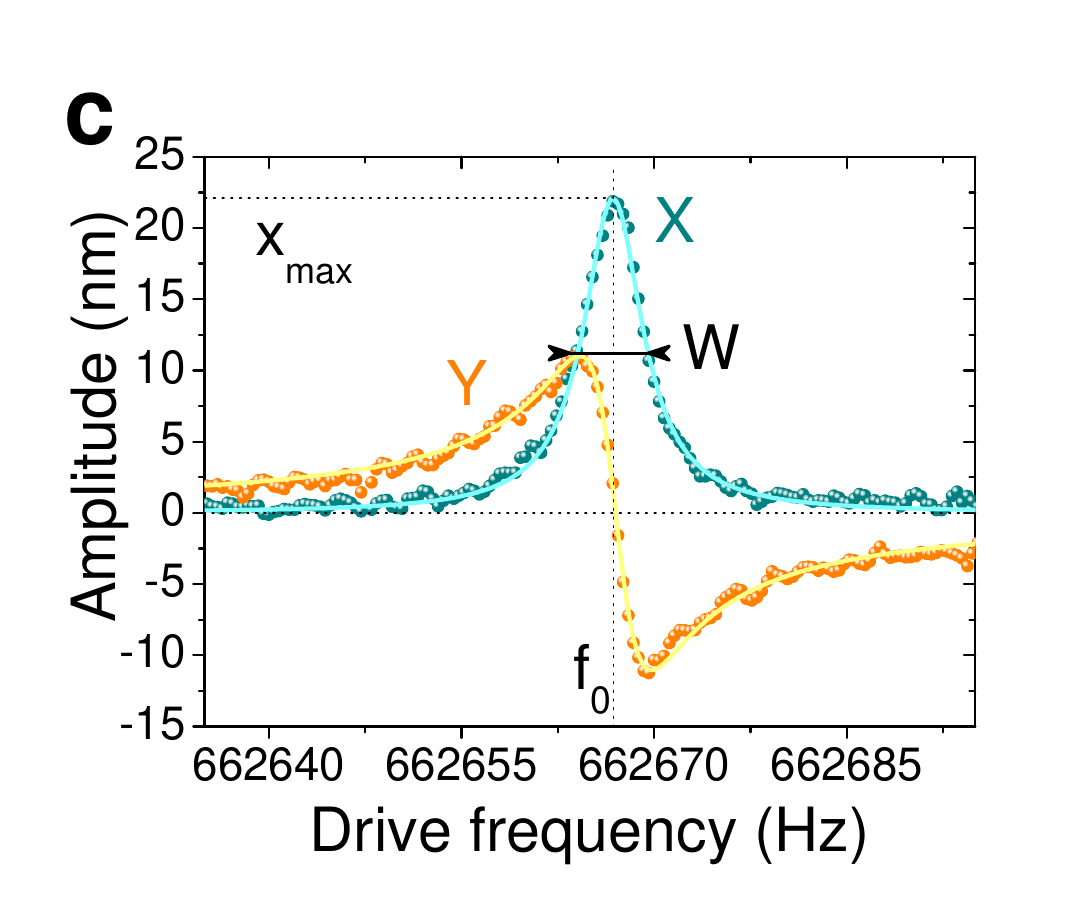}	
		 	 \end{minipage}			
		 	 	\begin{minipage}[c]{.45\linewidth}		
			 \includegraphics[scale=0.35]{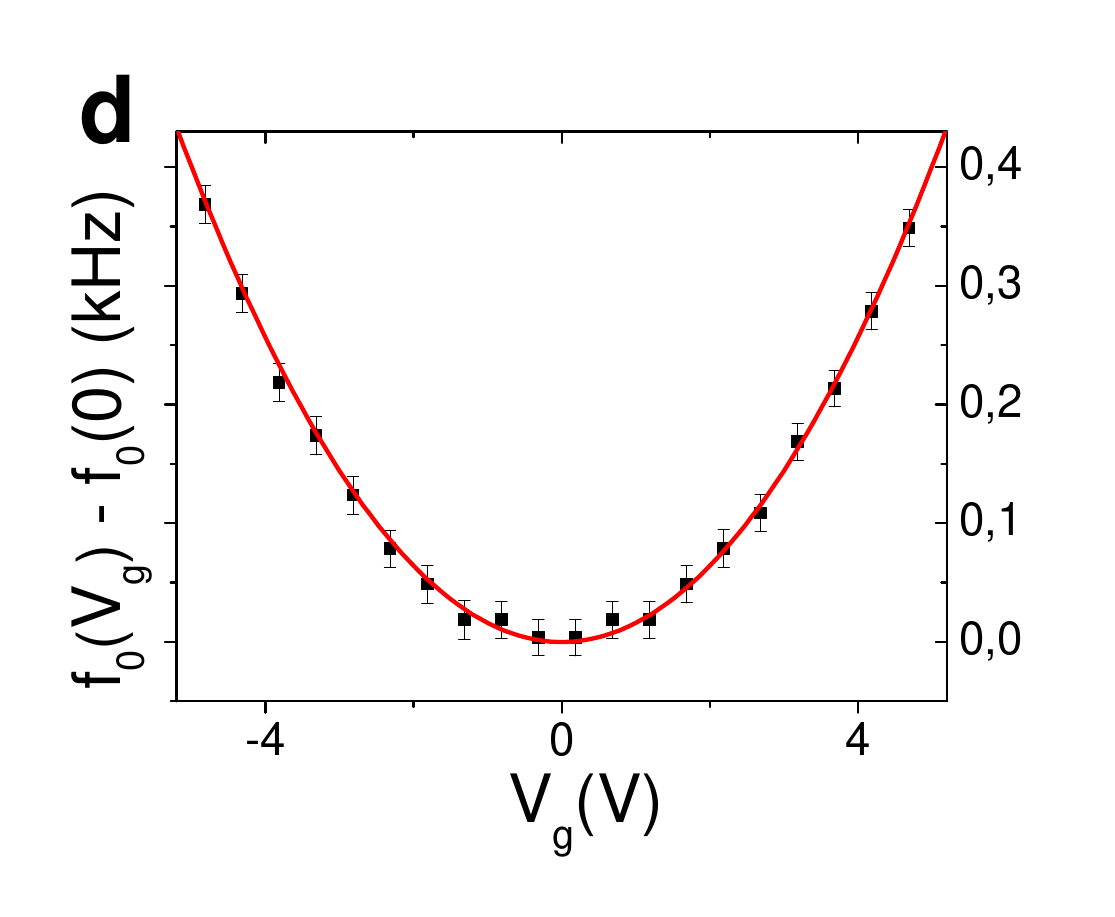}	 	
			 \end{minipage}	
			\caption{(Color online) 
			(a) Schematic view of the measurement setup. A sinusoidal current is fed into the cryogenic NEMS, actuated and detected through the magnetomotive technique. The detected signal is processed with a lock-in amplifier. An arbitrary waveform generator delivers the gate voltage. The bandpass filter is used only when the injected voltage is a Gaussian noise (typical histogram shown, with fit leading to $\sigma_	V$, see Appendix \ref{AppendixProba}). (b) Coloured SEM picture of the nanomechanical oscillator capacitively coupled to a gate electrode. (c) Standard linear response to a sinusoidal driving force of amplitude $F_0=83~\femto\newton$, with its in-phase $\mathrm{X}(\omega)$ and quadrature $\mathrm{Y}(\omega)$ components. Solid lines correspond to a (complex) Lorentzian line shape fit, with full width at half maximum (FWHM) $W$, position $f_0=\omega_0/(2\pi)$ and height $x_{max}$. (d) Shift of the NEMS resonance frequency $f_0$ as a DC voltage is applied to the gate electrode. The red solid line is a parabolic fit with a coefficient $\alpha/(2\pi)$.}
			\label{fig:setup}
		\end{figure}

Fig. \ref{fig:setup}b) shows a scanning electron microscope picture of the measured NEMS. This device consists in a $295~\micro\meter$ long x $100~\nano\meter$ thick x $300~\nano\meter$ wide doubly-clamped high-stress silicon nitride (Si$_3$N$_4$) suspended beam, with a $90~\nano\meter$ thick conducting aluminium layer on top. A gate electrode is made in the vicinity of the beam, separated by a gap $g\approx 3~\micro\meter$, over a $250~\micro\meter$ length. The total mass of the device is $M=4.7\times 10^{-14}~\kilo\gram$. The experiment is performed at $4.2~\kelvin$ in a cryogenic vacuum (pressure $<10^{-6}~\milli\bbar$).	

We first perform a careful calibration of the whole setup following Ref.\cite{RSICollin}, giving access to displacements and injected driving force in real units. Fig. \ref{fig:setup}a) shows the schematic of the experimental setup. A drive voltage is applied through a $1~\kilo\ohm$ bias resistor, injecting a current in the aluminium layer of the NEMS. Actuation and detection follow the magnetomotive scheme \cite{RSICollin,APLCleland}, with a magnetic field $B<1~\tesla$. The voltage induced by the motion is measured by standard lock-in detection with its in-phase (X) and quadrature (Y) components. 

We work on the fundamental flexural mode $n=0$, having effective spring constant $k_0$, mass $m_0$ and applied sinusoidal force amplitude $F_0$.						

The typical response to a sinusoidal driving force in the linear regime is displayed in Fig. \ref{fig:setup}c). The resonance frequency is measured at $\omega_0=2\pi\times 0.66~\mega\hertz$, which is in good agreement with calculated properties. In the following, in order to minimize electrical losses \cite{SensActCleland}, we keep a low magnetic field $B=0.1~\tesla$. The damping rate is then $\Delta\omega_0=2\pi\times 5.6~\hertz$, hence a quality factor $Q\approx 118~000$. Here $Q\gg 1$ so we can write the susceptibility in the standard Lorentzian approximation: $\chi_0(\omega)=\left[2m_0\omega_0(\omega_0-\omega-i\Delta\omega_0/2)\right]^{-1}$.
In our notations, we define in-phase $\mathrm{X}(\omega)=\mathrm{Im}[\chi_0(\omega)]F_0$ and quadrature $\mathrm{Y}(\omega)=\mathrm{Re}[\chi_0(\omega)]F_0$ components of the motion.
%

%
%
%
%
		\begin{figure*}[t!]		 
			 \includegraphics[scale=1.67]{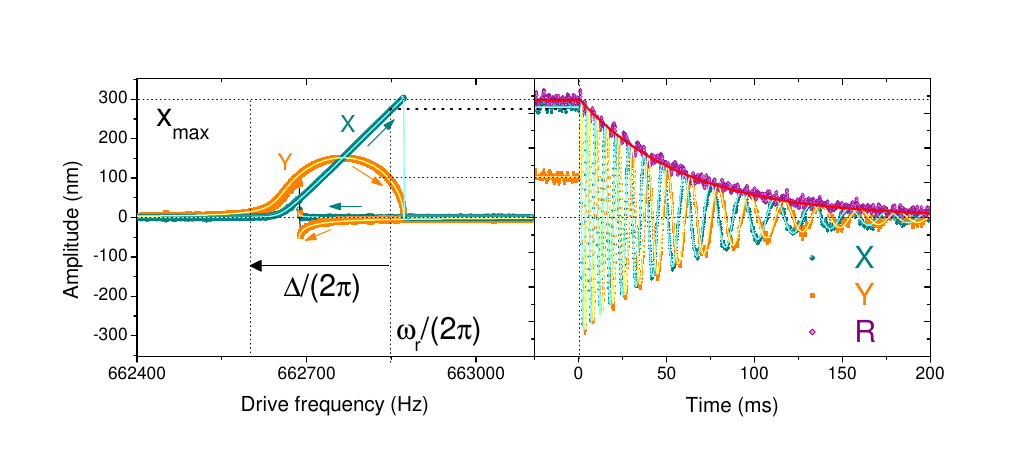}	
			\caption{\small{(Color online) Frequency and time domain response of the mechanical oscillator, in $\mathrm{X}$ and $\mathrm{Y}$ quadratures. The applied sinusoidal driving force of amplitude $F_0=1.1~\pico\newton$ enables the resonator to reach the bistable regime. The two branches of the hysteretic response are displayed in the frequency domain on the left panel (see arrows for the frequency sweep direction). The free decay of the steady state response (right panel) is measured after detuning the drive from the resonance $\omega_r$ by $\Delta=-2\pi\times 250~\hertz$. Solid lines are theoretical fits from the Duffing oscillator model (left) and the Lindstedt-Poincar\'e method (right). The magenta curve on the right panel corresponds to the amplitude $\mathrm{R}=\sqrt{\mathrm{X}^2+\mathrm{Y}^2}$, fit to an exponential decay of time constant $T_1$ (full curve).}}
			\label{fig:fitNL}
		\end{figure*}

A DC voltage $V_g$ is then delivered with an arbitrary waveform generator (AWG) on the gate electrode, which forms a geometric capacitance $C(x)$ with the beam itself. The energy stored by the capacitor can be written as $E_C=C(x)V_g^2/2$. Thus the additional electrostatic force acting on the beam along the $x$ axis is $F_C=(V_g^2/2)\partial C(x)/\partial x$. Expanding $F_C$ in a Taylor series \cite{RSICollin,APLKozinsky}, we obtain the effect of the DC voltage on the resonance through a frequency pulling term:
\begin{equation}
	\label{freqtuning}
\omega_0(V_g)-\omega_0(0)\approx-\frac{1}{4m_0\omega_0}\frac{\partial^2C(0)}{\partial x^2}V_g^2=\alpha V_g^2
	\end{equation}
Here $\partial^2C(0)/\partial x^2$ is defined as the coupling strength, and $\alpha>0$ is a constant only determined by the oscillator's intrinsic quantities and the coupling strength. We measure $\alpha=2\pi\times 16.44~\hertz/\volt^2$ [see Fig. \ref{fig:setup} d)], hence a coupling strength $\partial^2C(0)/\partial x^2=-4.3\times 10^{-5}~\farad/\meter^2$.
The voltage $V_g$ also generates a static force on the NEMS through the first order term of the $F_C$ expansion, namely $\frac{1}{2}\frac{\partial C(0)}{\partial x} V_g^2$. This drive is non-resonant with the mode under study, and has thus no impact on the dynamics discussed in the present article.
Higher-order terms contribute to the nonlinear (Duffing) coefficient \cite{RSICollin}. They can be safely neglected here since in Sec. \ref{dephaspart}, we keep the drive low enough to remain in the linear (Lorentzian-shaped) regime [Fig. \ref{fig:setup} c].
 More details on fabrication, the magnetomotive scheme, setup and capacitive coupling can be found in Supplementary Material \cite{SI}.
		
\section{DAMPING AND DEPHASING IN THE NONLINEAR REGIME}
\label{nonlinpart}
		
When driven with high input forces, our NEMS' oscillation amplitude at resonance becomes high enough so its dynamics enters the so-called "Duffing" regime, arising from a geometric non-linearity: the tensioning effect. The frequency domain response bends toward higher frequency and the resonance is shifted quadratically with the motion, eventually leading to a bistable regime \cite{NatCommsVenstra,Yamaguchi}. In doubly-clamped SiN nano-beams this non-linearity is known to be important \cite{NLOscNayfeh}, which means that already modest drive excitations can bring the device into the Duffing regime. This fact naturally brings in two questions: are there as well any nonlinear relaxation/decoherence processes to discover 
 in SiN structures, like for nanotubes \cite{NatCommsSchneider,NatNanoEichler} ? And how do we perform $T_1$ and $T_2$ measurements within a highly nonlinear/bistable dynamic system ?

In the following we describe how to measure $T_1$ and $T_2$ response times that we defined in analogy with NMR over the whole dynamic range, in order to compare them as a function of motion amplitude. The relaxation time $T_1$ obtained from ringdown measurements is only sensitive to dissipation mechanisms, by analogy with longitudinal relaxation time in the Bloch sphere for two-level systems. The ringdown is triggered by exciting the device at large motion amplitude at frequency $\omega_r$ close to $\omega_0$, and suddenly detuning the driving force to a frequency $\omega_r+\Delta$ where the device is off resonance. The beating of the mechanics with the local oscillator is then recorded \cite{PRBCollin} (see Fig. \ref{fig:fitNL} right panel). To obtain the plots, about $N\simeq 100$ to $1000$ decays have been averaged. The $T_2$ time is obtained from frequency sweeps across the resonance, with each point averaged over a time long enough to make sure that we are sensitive to potential low-frequency fluctuations (see Fig. \ref{fig:fitNL} left panel). $T_2$ is affected both by dissipation and phase fluctuations, in a similar manner to the transverse relaxation time in NMR or qubit measurements. But one fundamental difference is that here we are dealing with an almost harmonic classical oscillator, not a quantum two-level system, so the analogy is restricted to the phenomenology of the decoherence phenomenon. Moreover, while for qubits and spins the pure dephasing $\Gamma_{\phi}$ is defined through the comparison between $T_2$ and $2T_1$ \cite{PRBIthier}, in our system $T_2$ has to be compared to $T_1$, similarly to the case of a classical spin \cite{NatPhysFaust}. 
 \vspace*{1mm}

\begin{figure}		 
			 \includegraphics[width=8cm]{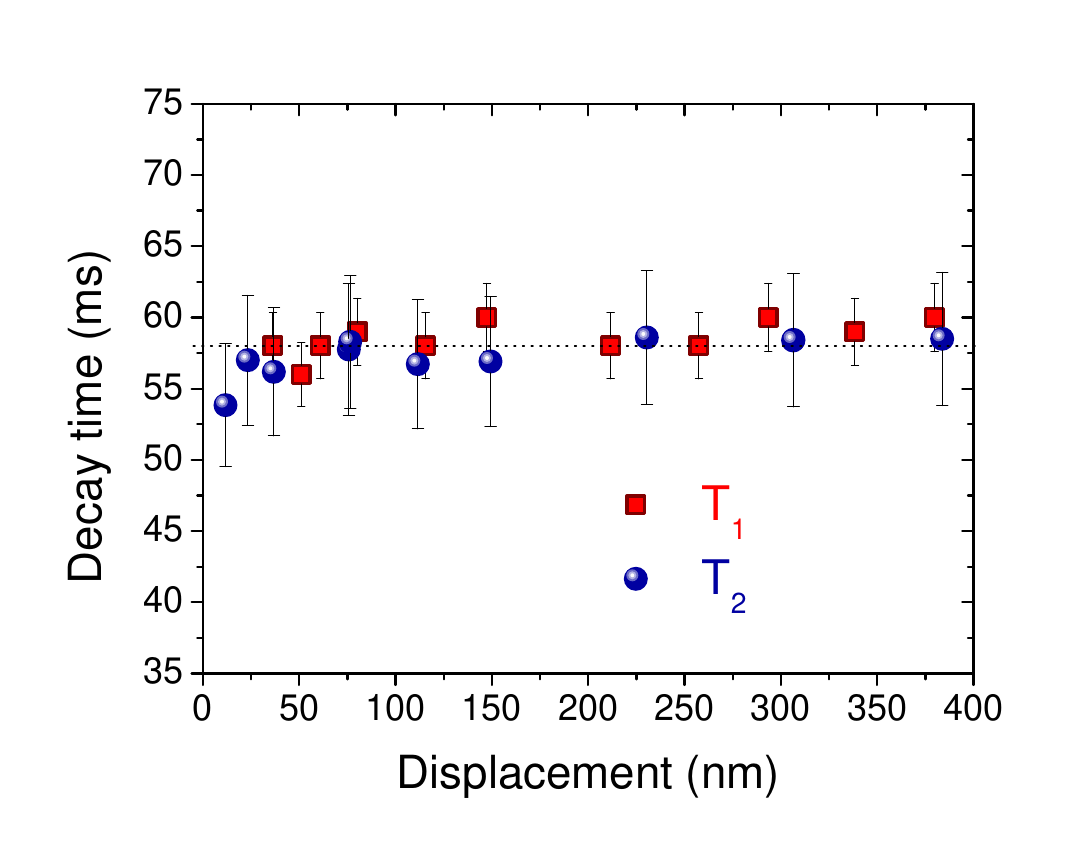}			
			\caption{\small{(Color online) Relaxation ($T_1$) and decoherence ($T_2$) times as a function of motion amplitude. The dotted line is set at $58~\milli\second$.}}
			\label{fig:T1T2NL}
		\end{figure}

Fig. \ref{fig:fitNL} shows both frequency and time-domain measurements in a highly nonlinear regime, with a peak-to-peak maximum amplitude of about $600~\nano\meter$, that is, more than three times the NEMS thickness. Here the bistable resonance line can be fit according to the Duffing model for nonlinear oscillators \cite{PRBCollin}. In the linear regime, the FWHM $W$ of the X quadrature directly yields the Lorentzian parameter $\Delta\omega_0/(2\pi)$ and is used to define $T_2=(\pi W)^{-1}$. In the strongly nonlinear case, from the measured maximum amplitude $x_{max}$ one can calculate \cite{PRBCollin} the intrinsic linewidth parameter $\Delta\omega_0$ used for the full nonlinear fit: $\Delta\omega_0=F_0/(m_0\omega_0x_{max})$. This parameter can again be converted into a $T_2$. The latter is displayed as a function of motion amplitude in Fig. \ref{fig:T1T2NL} (blue dots). The time-decay oscillation can be fit using the Lindstedt-Poincar\'e method \cite{PRBCollin}. The relaxation is found to be exponential over the entire range studied, leading to a $T_1$ showed in Fig. \ref{fig:T1T2NL} (red dots). \\

The ringdown measurement yields a relaxation time $T_1\simeq 58~\milli\second\pm 3~\milli\second$ in the whole attainable amplitude range, which matches the $T_2$ time of spectral data defined as $2/\Delta\omega_0\simeq 57~\milli\second\pm 3~\milli\second$: we thus verify $T_1=T_2$ within $\pm~8~\%$. In other words, we do not identify any sources of nonlinear damping or dephasing in our SiN system, as opposed to carbon-based devices \cite{NatCommsSchneider,NatNanoEichler}. This result is achieved over an unprecendently broad dynamic range, with a measurement technique that can be adapted to other devices, in particular graphene or nanotube ones. As far as the present work is concerned, this establishes our device as an ideal NEMS resonator.

\section{CONTROLLING PURE DEPHASING}
\label{dephaspart}

In order to introduce a dephasing process, we deliver a Gaussian noise with the AWG connected to the gate electrode, instead of the DC voltage used previously for calibration purposes. Our aim is to model "slow" fluctuations of the resonance frequency, that is with correlation times longer than the relaxation time $T_1$ \cite{PRLZhang}. In the aforementioned work, the interplay of the driving force with the frequency noise spectrum is characterized, in order to separate fast and slow noise contributions. Complementarily, we focus here on how the driven response evolves with respect to the level of slow fluctuations. Furthermore, in Ref. \cite{PRLZhang}, the frequency noise is experimentally taken to be Gaussian; we on the other hand extend the study to highly asymmetric statistics. 
The modelling is fairly generic (see Appendix \ref{AppendixConvolutPic}), and does apply as well to the case of slow Telegraph frequency noise \cite{Telegraph}.  
%
%
%
%
%
%

\begin{figure}		 
			 \includegraphics[width=8cm]{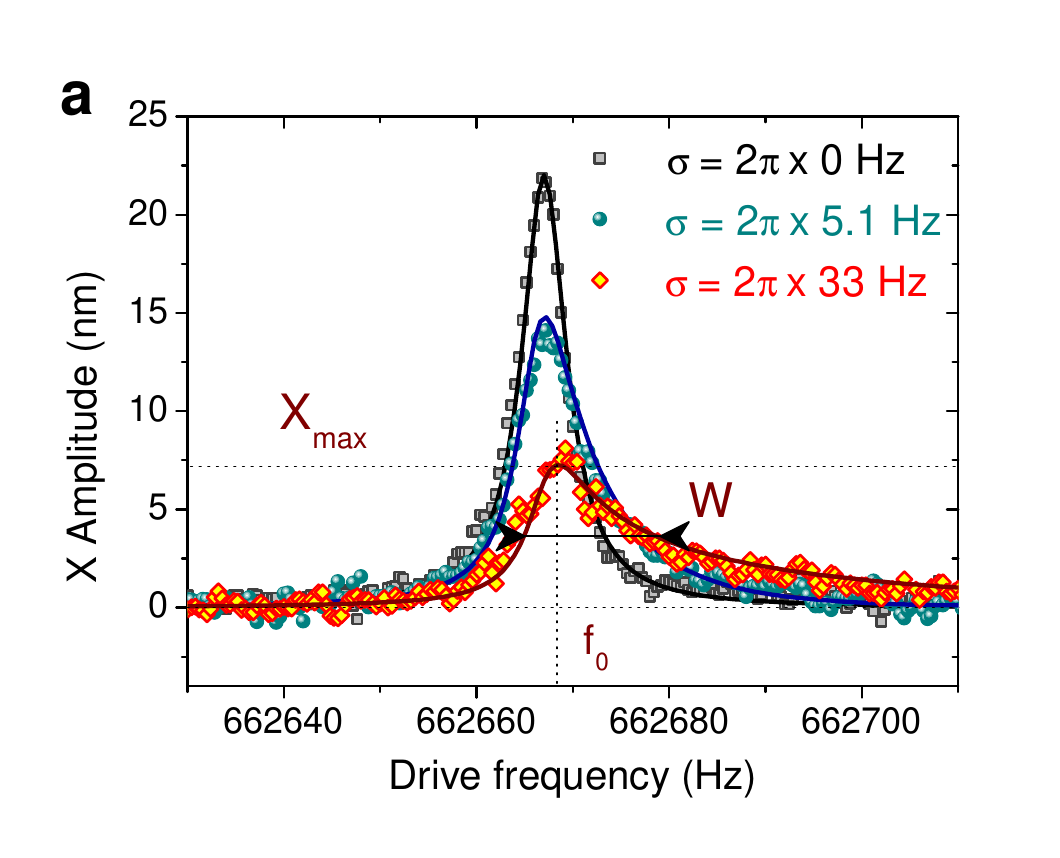}			 			 \includegraphics[width=8cm]{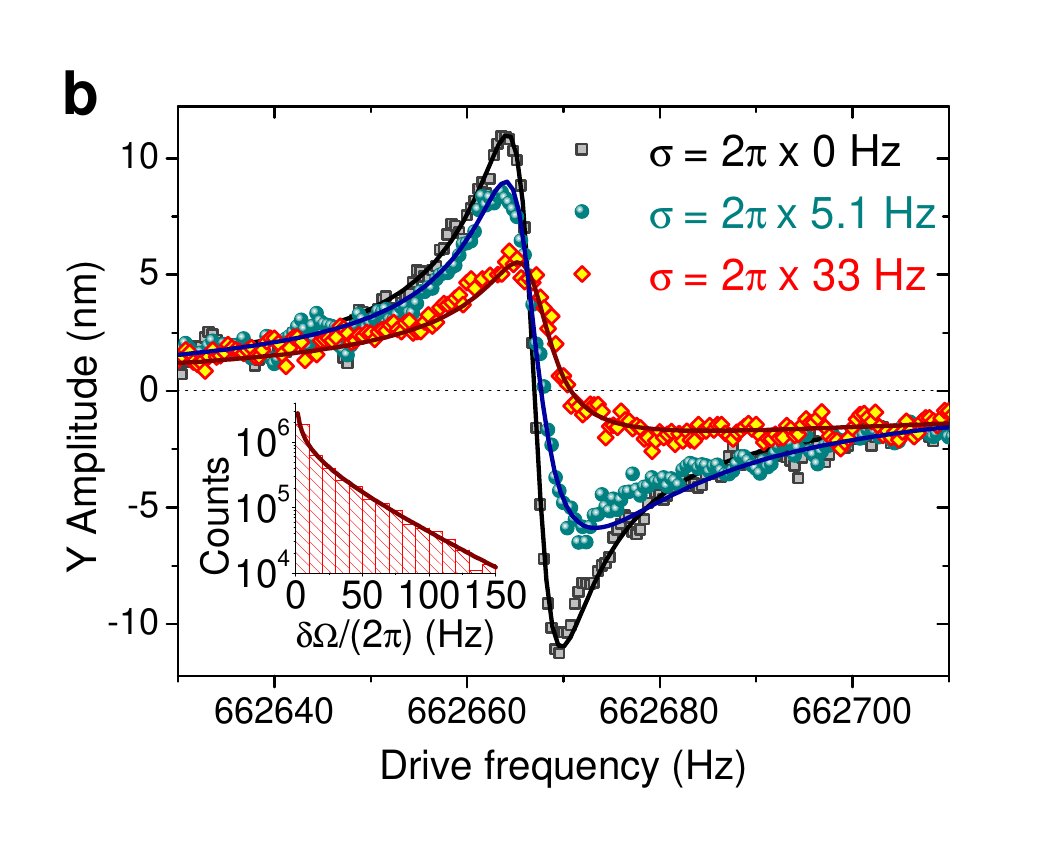}
			\caption{(Color online) Frequency sweep measurements in (a) phase (X) and (b) quadrature (Y) components of the mechanical response $x_{\sigma}(\omega)=\chi_{\sigma}(\omega)F_0$, with a sinusoidal driving force of amplitude $F_0=83~\femto\newton$. Solid lines are theoretical curves calculated from Eq. (\ref{chi_alt}), and resonance line parameters $W, f_0$ are shown in a). Inset in (b): frequency noise histogram at $\sigma=2\pi\times 33~\hertz$ calculated from $\delta V_g^2$ measurements, for $4\times 10^6$ counts. The solid line is the application of Eq. (\ref{distribution}).}
			\label{fig:altLor}
		\end{figure}

To keep only low-frequency fluctuations while avoiding spurious $1/f$ noise coming from the setup, the delivered voltage noise is thus filtered [see Fig. \ref{fig:setup}a)] below $40~\milli\hertz$ and above $0.8~\hertz$ ($-3$ dB cut-off frequencies). The resulting voltage fluctuations on the gate electrode, represented by the Gaussian random variable $\delta V_g$, of mean value $0$ and standard deviation value $\sigma_V=\langle\delta V_g^2\rangle^{1/2}$, translate through the capacitive coupling into fluctuations of the resonance frequency. Following Eq. (\ref{freqtuning}) the new random variable associated with these fluctuations is $\delta\Omega(t)=\alpha\delta V_g^2(t)$, with a probability distribution 
[see Fig. \ref{fig:altLor}b) inset]:		
	\begin{equation}
	\label{distribution}
\rho_{\sigma}(\delta\Omega)=\theta(\delta\Omega) \left(\frac{1}{\sqrt{2}\pi\sigma\delta\Omega}\right)^{1/2}\exp\left(-\frac{\delta\Omega}{\sqrt{2}\sigma}\right)
	\end{equation}	
where $\theta(\delta\Omega)$ is the Heaviside step function, and $\sigma=\sqrt{2}\alpha\sigma_V^2=[\langle\delta\Omega^2\rangle-\langle\delta\Omega\rangle^2]^{1/2}$ is the standard deviation of the distribution, taken as a characteristic frequency noise level (in rad.s$^{-1}$). This is the effective tunable parameter with which we characterize decoherence in our system. 
Mathematically, we show that the averaging of fluctuations can be treated similarly to the case of "inhomogeneous broadening" in NMR \cite{abragam}, spatial-disorder being replaced, as for qubits, by time-dependent disorder (see Appendix \ref{AppendixConvolutPic}). 
In this context, the $T_2$ time is re-named $T_2^*$. However for simplicity we keep here the notation without the star, since our aim is to model the impact of {\it any } decoherence mechanism, especially intrinsic ones. Even though our noise source is extrinsic.
Knowing the distribution $\rho_{\sigma}$, the altered susceptibility can thus be written as the convolution of the mechanical response $\chi_0$ in the high Q Lorentzian limit by the distribution $\rho_{\sigma}$:
\begin{equation}
	\label{chi_alt}
\chi_{\sigma}(\omega)=\int_{-\infty}^{+\infty}\chi_0(\omega-\delta\Omega)\rho_{\sigma}(\delta\Omega)\mathrm{d\delta\Omega}
	\end{equation}
Note that $\sigma$ is also defined through $S_\Omega(\omega)$, the spectrum of the frequency fluctuations, with $\langle\delta\Omega^2\rangle=(2\pi)^{-1}\int S_\Omega(\omega)\mathrm{d}\omega=3\sigma^2/2$. Thus Eq. (\ref{chi_alt}) is valid only if the noise power is integrable (which justifies mathematically the introduction of a higher cut-off frequency). 
 In particular for $1/f$ noise, dephasing would depend on the measurement time since the lower bound of the spectrum would be fixed by the experiment duration itself (which justifies our lower cut-off filtering choice). 

Fig. \ref{fig:altLor} shows frequency domain measurements for different noise levels, with theoretical calculations without free parameters from Eq. (\ref{chi_alt}), clearly showing spectral broadening. The convolution picture provides a rather intuitive interpretation: while the  noise level $\sigma$ responsible for pure dephasing is kept small enough compared to the mechanical damping rate $\Delta\omega_0$, one can approximate the distribution $\rho_{\sigma}$ as a Dirac delta function, and then the convolution leaves the Lorentzian susceptibility unchanged. When $\sigma$ becomes comparable to the damping rate, the distribution leaves its imprint in the mechanical response, as an evidence for mechanical dephasing: here, the resonance line is asymmetrically broadened, which is a signature for the non-Gaussian nature of the fluctuations (see discussion below). For large fluctuations ($\sigma\gg\Delta\omega_0$) a significant deviation from the Lorentzian lineshape is observed (see discussion below) and the resonance peak looks like the distribution $\rho_{\sigma}$ itself, the initial susceptibility being almost a Delta function.

%
%
%
%
%
%
%

\begin{figure}		 
\includegraphics[scale=0.75]{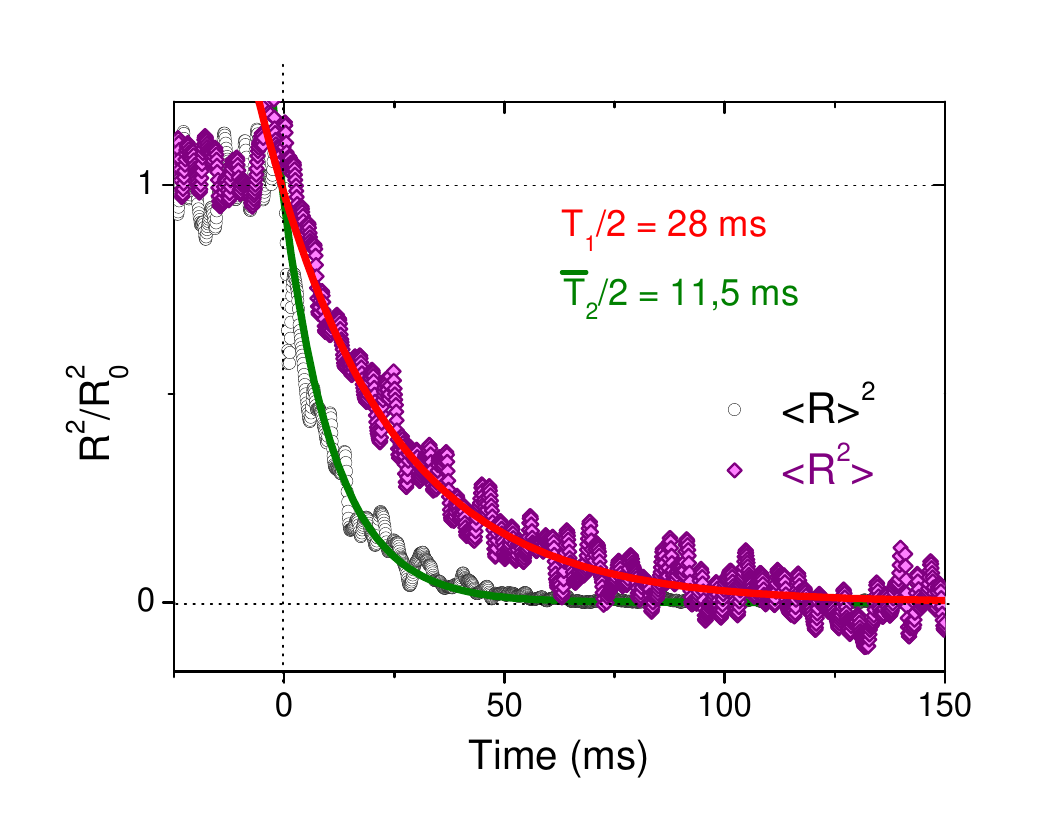}
\caption{(Color online) Time domain measurements of $\langle\mathrm{R}^2\rangle_N$ (magenta) and $\langle\mathrm{R}\rangle_N^2$ (black, empty dots), normalized to their steady state resonant amplitudes of motion $\langle\mathrm{R}_0^2\rangle_N$ and $\langle\mathrm{R}_0\rangle_N^2$. Solid lines are exponential fits. Data acquired for $\sigma=2\pi\times 33~\hertz, F_0=83~\femto\newton$.}
\label{fig:TimeMeasDiff}
\end{figure}

\begin{figure}[h!]		 
			 \includegraphics[width=8cm]{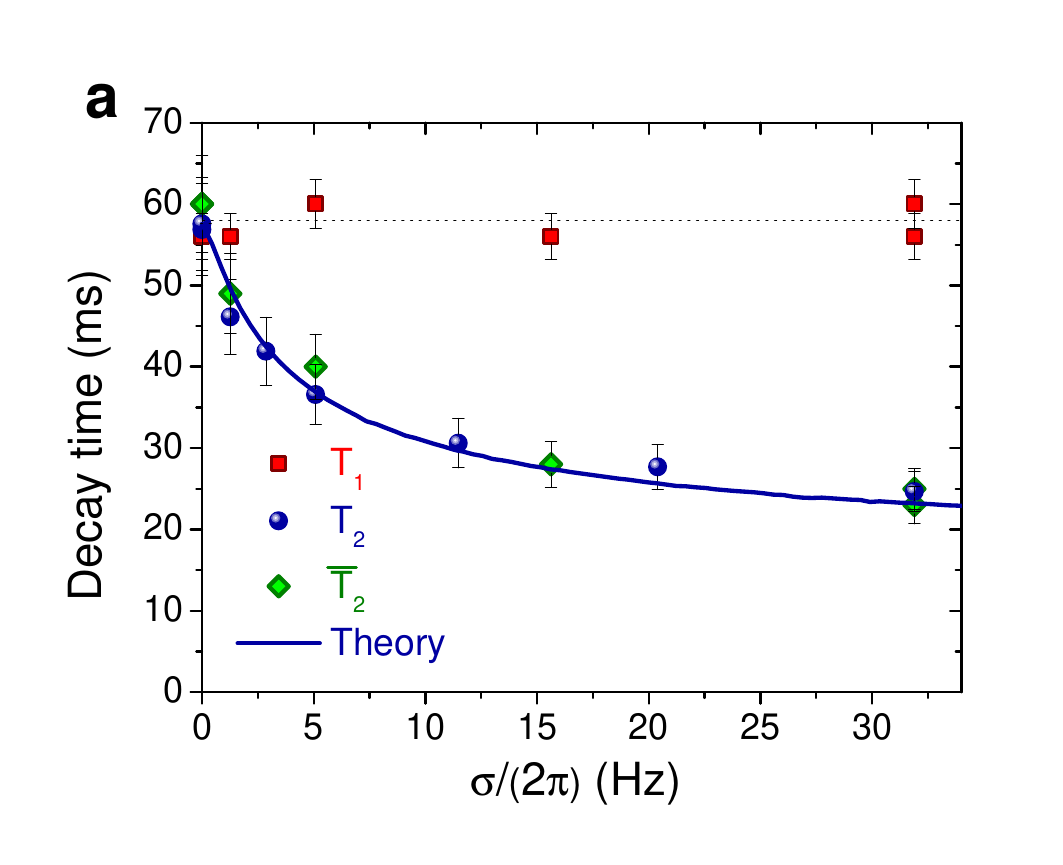}
			 \includegraphics[scale=0.72]{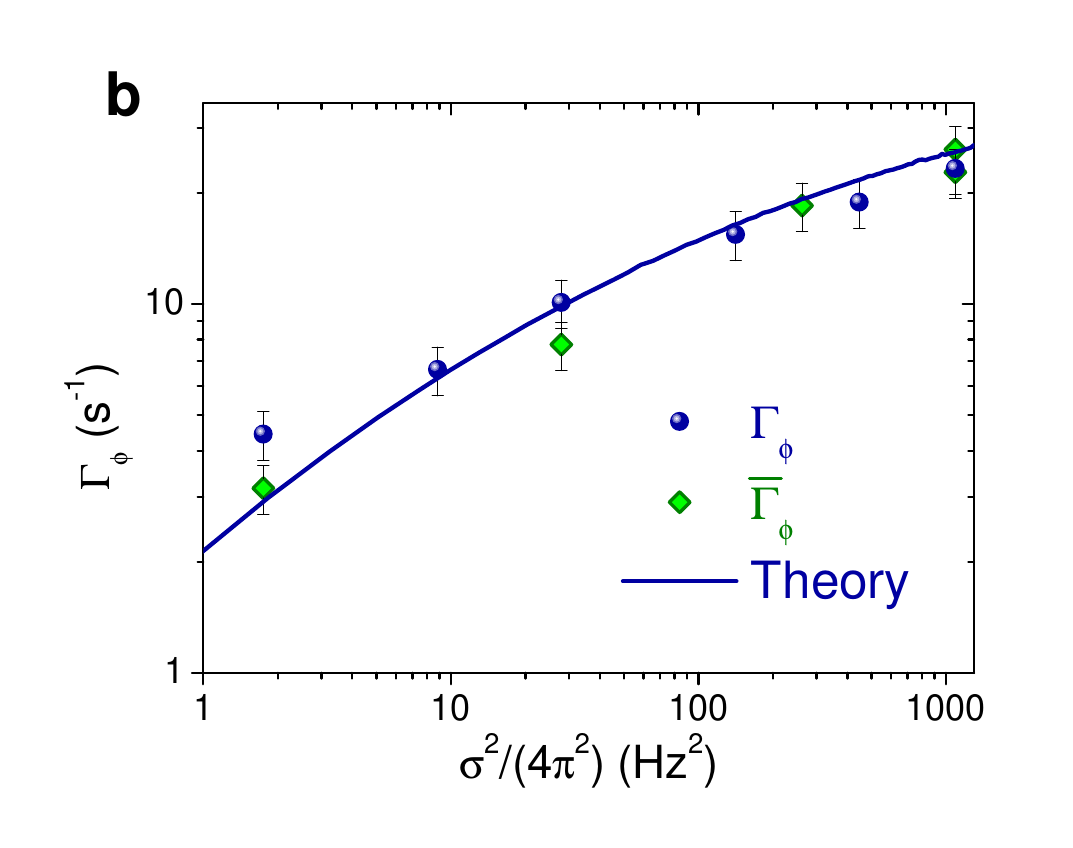}
\caption{(Color online) (a) Measured decay times as a function of injected frequency noise level $\sigma/(2\pi)$. The blue solid line is a theoretical calculation from Eq. (\ref{chi_alt}). Blue dots are obtained from frequency domain measurements, while green dots are obtained from exponential fits of $\langle\mathrm{R}\rangle_N^2$ in time domain measurements. Red dots are time domain measurements of energy relaxation, with the dotted line used as a guide for the eyes. (b) Pure dephasing rate as a function of frequency noise level. Blue dots account for frequency domain measurements, green dots represent data from time domain ($\overline{T}_2$) measurements. The blue solid line is the theoretical curve calculated from Eq. (\ref{chi_alt}).}
\label{fig:decoh}
\end{figure}

The time domain measurements (see Fig. \ref{fig:TimeMeasDiff}) performed for different noise levels yield unambiguously the same decay time $T_1\simeq 57~\milli\second\simeq 1/(\pi\times 5.6~\hertz)$ within $\pm~5\%$ for all data sets [see Fig. \ref{fig:TimeMeasDiff} and \ref{fig:decoh}a)]. Interestingly, $T_2$ can also be measured via time-domain measurements, by averaging first X and Y quadratures over $N$ relaxations, and then computing $\langle\mathrm{R}\rangle_N^2=\langle\mathrm{X}\rangle_N^2+\langle\mathrm{Y}\rangle_N^2$. Indeed the quadratures taken independently are sensitive to frequency fluctuations, while the direct measurement of $\langle\mathrm{R}^2\rangle_N=\langle\mathrm{X}^2+\mathrm{Y}^2\rangle_N$ is not. The obtained curve (see Fig. \ref{fig:TimeMeasDiff}) 
whose exact mathematical description is given in Appendix \ref{AppendixShape}, can be well fit experimentally with an exponential law 
yielding a decay time in very good agreement with the $T_2$ defined from frequency domain measurements [see Fig. \ref{fig:decoh}a)]. For simplicity we call this time domain measurement of decoherence $\overline{T}_2$.
For a quantitative discussion we plot in Fig. \ref{fig:decoh}a) the $T_2=(\pi W)^{-1}$ obtained from a simple estimate of the FWHM in the frequency domain [see Fig. \ref{fig:altLor}a)], together with $\overline{T}_2$ and $T_1$. Starting at the initial value of $2/\Delta\omega_0\approx 57~\milli\second$ at $\sigma=0$, $T_2$ decreases to less than half its noise-free value for $\sigma=2\pi\times 33~\hertz$. Our data are in good agreement (within $\pm~10\%$) with the theoretical curve calculated from the resonance lines derived from Eq. (\ref{chi_alt}), with no free parameters. 

Clearly, as soon as a significant frequency noise is applied, we verify $T_1\neq T_2$, thus demonstrating the impact of artificially engineered dephasing. This can also be seen as a consequence of the dispersive coupling of the mechanics to fluctuations described by the convolution formalism: the area under the X quadrature resonance curve, which corresponds to the energy stored in the mechanical mode, is preserved (more mathematical details in Appendix \ref{AppendixConvolutPic}).

One can then extract the pure dephasing rate $\Gamma_{\phi}=T_2^{-1}-T_1^{-1}$, as displayed on Fig. \ref{fig:decoh}b). The theoretical curve fits within $\pm~15\%$ the data, demonstrating good agreement. This final graph thus presents the relation between $\Gamma_{\phi}$ and $\sigma^2\propto\int S_{\Omega}(\omega)\mathrm{d}\omega$, which is the classical mechanical analogue to the superconducting qubit expression which  can be found in Ref. \cite{PRBIthier}. Note that the shape of the curve in Fig. \ref{fig:decoh}b) is also characteristic of the frequency noise probability distribution.
The mathematical tools are described thoroughly in Appendix \ref{AppendixConvolutPic}, and can be easily applied as a versatile method describing decoherence even in complicated non-analytic cases like the one presented here. 

While being a practical quantitative estimate of decoherence, the definition of  $T_2=(\pi W)^{-1}$ or the exponential fit of $\overline{T}_2$ essentially amount to approximating the resonance lineshapes by Lorentzians, which is rather crude. However, defining a shape factor $S=2\pi W x_{max}/\int\mathrm{Im}[\chi_0(\omega)]F_0\mathrm{d}\omega$, one can verify how accurate this simple approximation is. For the initial Lorentzian susceptibility we obtain $S=2/\pi$ as we should. For our strongest level of noise, $S$ only deviates by $20\%$ from this value, which proves that the simple approach is reasonably accurate (see Appendix \ref{AppendixShape}).

But obviously, there is information in the measured resonance lineshapes that deserves to be exploited \cite{AddDyk}. Highlighting this point is precisely the reason why we created a frequency noise that is non-Gaussian and asymmetric. 
The complete fits of the lineshapes prove that the theory can handle non-conventional noises. Furthermore, we note that the asymmetry in the resonance is a direct image of the noise properties, as frequency fluctuations here 
take only positive values. While demanding, the deconvolution of noise from the measured lineshape should be feasible, reconstructing the actual fluctuations distribution. This could prove to be extremely useful in actual experimental systems where dephasing is caused by intrinsic unknown mechanisms. One mathematical technique which has been proposed in the literature is to compute the moments of the frequency response \cite{PRBMaizelis}, an approach also found in NMR \cite{abragam}.

In summary, we establish that without external influence, pure dephasing is negligible in high quality SiN nanomechanical devices. 
We measure $T_1$ and $T_2$ over an unprecedently large dynamic range, ruling out any possible nonlinear dephasing/damping mechanisms, as opposed to carbon-based systems. We then engineer slow fluctuations of the beam's resonance frequency with an external tunable source, unravelling the signatures of pure dephasing. By analogy with quantum bits, we quantitatively link the dephasing rate to the characteristic level of noise. 
We develop the methods that can be applied to a wide range of devices and experiments studying pure dephasing originating from slow frequency fluctuations.
Finally, we demonstrate that the lineshapes of the mechanical resonances contain valuable information about the distribution of the frequency fluctuations, and discuss the possibility to extract this information for actual nanomechanical systems suffering from intrinsic sources of noise. 

\section*{ACKNOWLEDGEMENTS}

We thank J. Minet and C. Guttin for help in setting up the experiment, as well as T. Fournier, J.-F. Motte and T. Crozes for help in the device fabrication. We acknowledge support from the ERC CoG grant ULT-NEMS No. 647917, from MICROKELVIN, the EU FRP7 grant No. 228464, and from the French ANR grant QNM No. 0404 01.

\appendix
\section{PROBABILITY DISTRIBUTION FOR FREQUENCY FLUCTUATIONS}
\label{AppendixProba}

\begin{figure}
			 \includegraphics[scale=0.65]{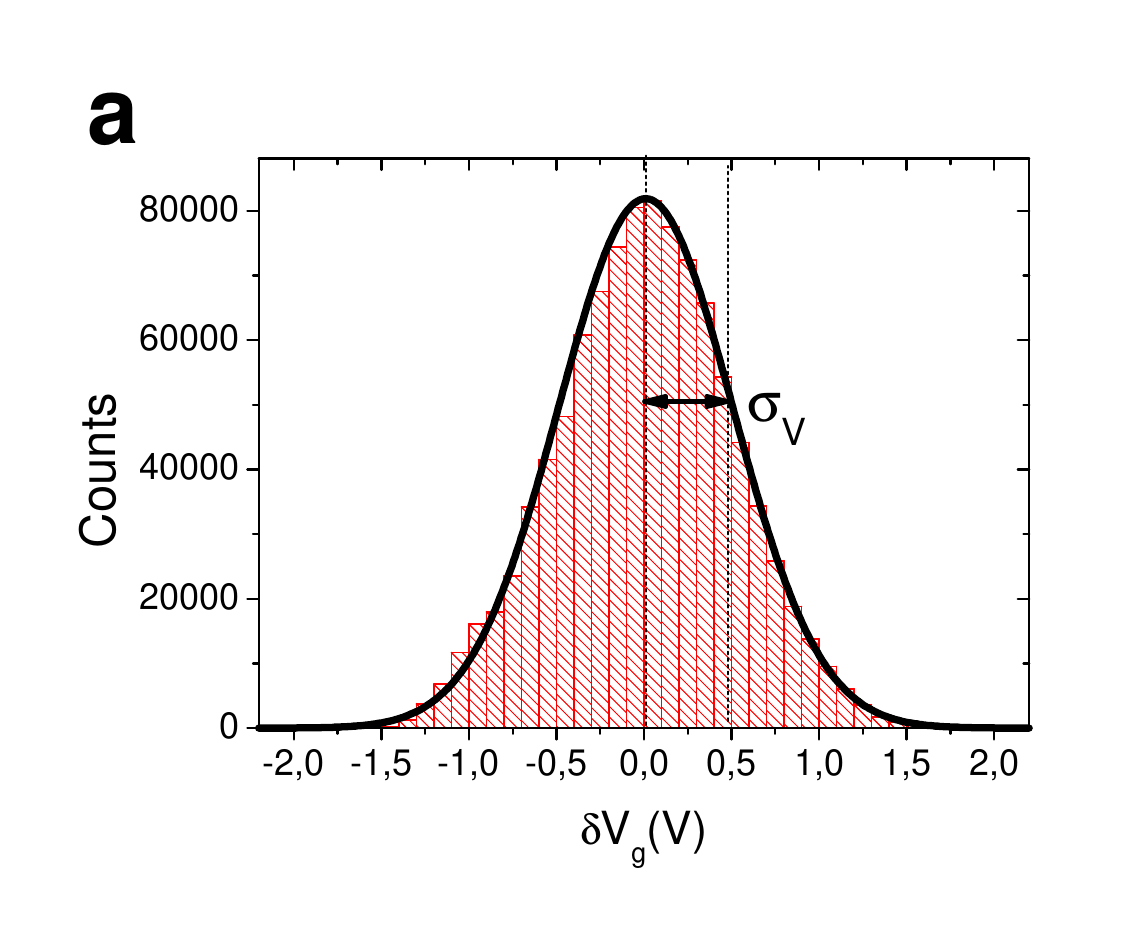}	
			  \includegraphics[scale=0.65]{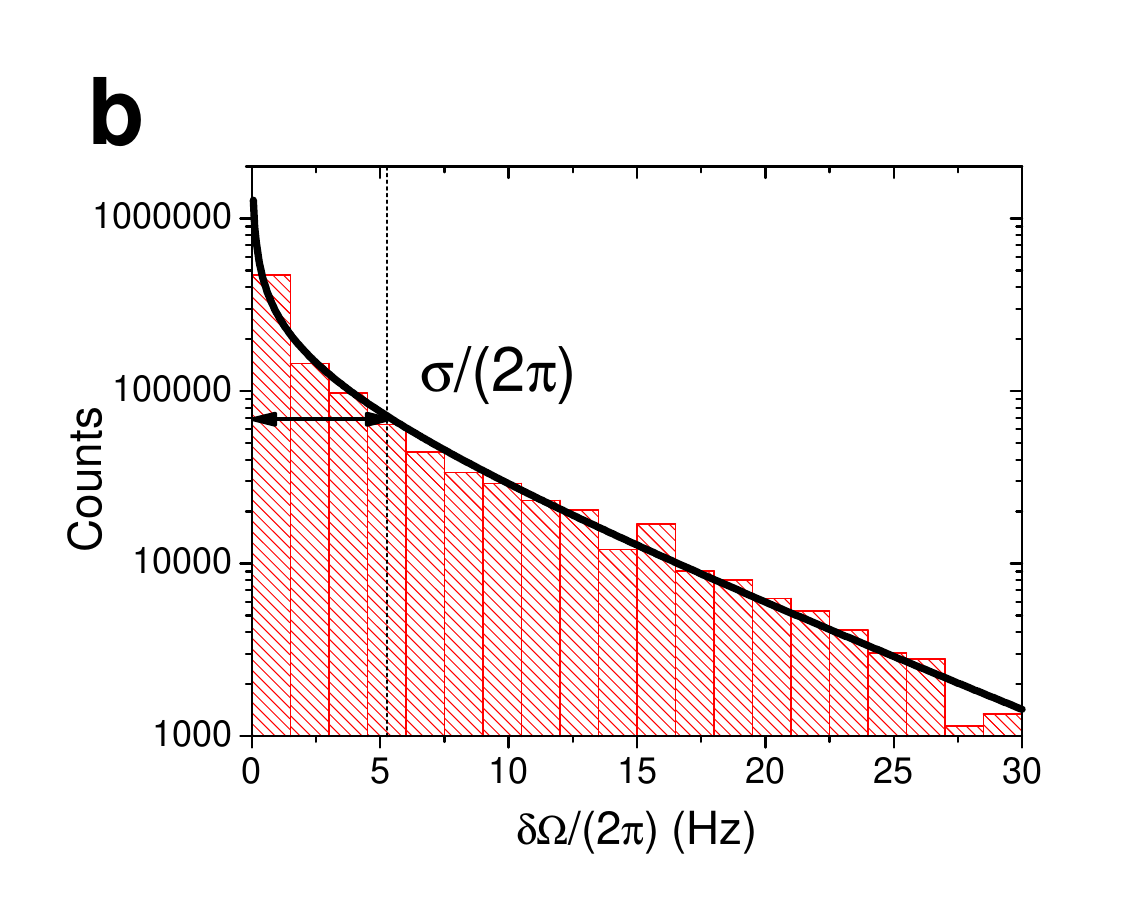} 
\caption{(Color online) (a) Gaussian distribution of a gate voltage noise injected with standard deviation value $\sigma_V=490~\milli\volt$, for $10^6$ samples. The solid line is the application of Eq. (\ref{volt_proba}). (b) Equivalent distribution for induced frequency fluctuations, of characteristic level $\sigma=2\pi\times 5.1~\hertz$ (note the log scale). The solid line is the application of Eq. (\ref{freq_proba}).}
\label{fig:distris}
\end{figure}

The voltage noise injected on the gate electrode is represented by a random Gaussian variable $\delta V_g(t)$ valued in $\mathbb{R}$, of mean value 0 and standard deviation value $\sigma_V=\left\langle \delta V_g^2\right\rangle^{1/2}$. Then, the probability to get a voltage valued in the interval $\left[\delta V_g, \delta V_g+\mathrm{d}\delta V_g\right]$ is:

\begin{equation}
\label{volt_proba}
\mathrm{d}p(\delta V_g\in\mathbb{R})=\frac{1}{\sigma_V\sqrt{2\pi}}\exp{\left(-\frac{\delta V_g^2}{2\sigma_V^2}\right)}\mathrm{d}\delta V_g.
\end{equation}

By capacitive coupling, these voltage fluctuations are converted to frequency fluctuations on the NEMS' resonance frequency, modelled by a random variable $\delta\Omega(t)=\alpha \delta V_g^2(t)$ which is by construction only valued in $\mathbb{R}^+$. By simple change of variable, the probability to get a frequency shift valued in the interval $\left[\delta\Omega,\delta\Omega+\mathrm{d}\delta\Omega\right]$ is:

\begin{equation}
\label{freq_proba}
\mathrm{d}\tilde{p}(\delta\Omega\in\mathbb{R}^+)=\sqrt{\frac{1}{\sqrt{2}\pi\sigma\delta\Omega}}\exp{\left(-\frac{\delta\Omega}{\sqrt{2}\sigma}\right)}\mathrm{d}\delta\Omega.
\end{equation}

One can easily see that this probability distribution is normalized. It can be proved using the normalization for the probability distribution in $\delta V_g$, which is symmetric around $0$ so we can write:

\begin{align*}
\label{normalization}
\int_{-\infty}^{+\infty}\mathrm{d}p(\delta V_g) &= 2\int_0^{+\infty}\frac{1}{\sigma_V\sqrt{2\pi}}\exp{\left(-\frac{\delta V_g^2}{2\sigma_V^2}\right)}\mathrm{d}\delta V_g\\
&= \int_{-\infty}^{+\infty}\rho_{\sigma}(\delta\Omega)\mathrm{d}\delta\Omega=1,
\end{align*}
hence the expression obtained for the probability distribution for frequency fluctuations $\rho_{\sigma}$ in the main article, Eq. (\ref{distribution}). 
The probability distributions are presented in Fig. \ref{fig:distris}.
Here $\sigma=\sqrt{2}\alpha\sigma_V^2$ is by construction the standard deviation of the frequency noise. Indeed, one can analytically obtain $M_n[\rho_{\sigma}]$, the $n^{\mathrm{th}}$ moment of the distribution $\rho_{\sigma}$:

\begin{equation}
\label{moments}
\begin{split}
M_n[\rho_{\sigma}]&=\int_{-\infty}^{+\infty}\delta\Omega^n\rho_{\sigma}(\delta\Omega)\mathrm{d}\delta\Omega\\&=\frac{(2n)!}{n!}\left(\frac{\sigma}{2\sqrt{2}}\right)^n.
\end{split}
\end{equation}

The present work aims at presenting a framework as generic as possible. This is why the distribution $\rho_{\sigma}$ is chosen non-Gaussian, and non-centered (its mean is {\it not} zero). We therefore deliberately keep this aspect in our discussion, and do not re-center the distribution in the following. 
Then, using the canonical definition of the squared standard deviation value of $\delta\Omega$:

\begin{equation}
\begin{split}
\label{RMSomega}
\langle\delta\Omega^2\rangle-\langle\delta\Omega\rangle^2&=M_2[\rho_{\sigma}]-M_1[\rho_{\sigma}]^2\\
&=\frac{3\sigma^2}{2}-\left(\frac{\sigma}{\sqrt{2}}\right)^2\\&=\sigma^2.
\end{split}
\end{equation}

Besides, the $2^{\mathrm{nd}}$ and $4^{\mathrm{th}}$ moments of the $\delta V_g$ Gaussian distribution are respectively $\sigma_V^2$ and $3\sigma_V^4$. Thus, for the sake of consistency with the experimental definition, the standard deviation can also be calculated as follows:

\begin{align*}
\sigma^2 &= \langle\delta\Omega^2\rangle-\langle\delta\Omega\rangle^2\\
			    &  
			    =2\alpha^2 \sigma_V^4.
\end{align*}

Note that all the moments of the distribution $\rho_{\sigma}$ are finite, making it suited to our study. 
The frequency fluctuations power spectral density $S_{\Omega}(\omega)$ reads, following Wiener-Khinchin relation:

\begin{equation}
\label{Wiener-Khinchin}
S_{\Omega}(\omega)=\int_{-\infty}^{+\infty}\langle\delta\Omega(t)\delta\Omega(t+\tau)\rangle\mathrm{e}^{-i\omega\tau}\mathrm{d}\tau.
\end{equation}

One gets by swapping the integrals:

\begin{equation}
\begin{split}
&\frac{1}{2\pi}\int_{-\infty}^{+\infty}S_{\Omega}(\omega)\mathrm{d}\omega\\
&=\langle\delta\Omega^2(t)\rangle=\frac{3\sigma^2}{2}.
\end{split}
\end{equation}

Thus the integrated value of the spectrum is related quadratically to the quantity $\sigma$.

\section{CONVOLUTION PICTURE}
\label{AppendixConvolutPic}

The susceptibility $\chi_0$ is randomized by the fluctuating gate voltage $\delta V_g$. It can be recasted from Newton's law as:

\begin{align*}
\label{RandomSusceptibilityVolt}
\chi_0(\omega,\delta V_g) &= \frac{1}{m_0\left[\omega_0^2-\frac{1}{2m_0}\frac{\partial^2C(0)}{\partial x^2}\delta V_g^2-\omega^2+i\omega\Delta\omega_0\right]}\\
&= \frac{1}{m_0(\omega_0^2+2\omega_0\delta\Omega-\omega^2+i\omega\Delta\omega_0)}.
\end{align*}

Assuming the fluctuations are small compared to the characteristic frequency $\omega_0$, one can then write in the Lorentzian limit ($\omega_0^2-\omega^2\approx 2\omega_0 (\omega_0-\omega)$, $\omega\Delta\omega_0\approx \omega_0\Delta\omega_0$):

\begin{equation}
\begin{split}
\label{RandomSusceptibilityFreq}
\chi_0(\omega,\delta\Omega)&\approx \frac{1}{2m_0\omega_0(\omega_0+\delta\Omega-\omega+i\frac{\Delta\omega_0}{2})}\\&=\chi_0(\omega-\delta\Omega).
\end{split}
\end{equation}

The susceptibility depends directly on the random variable $\delta\Omega$. At a given frequency $\omega$ and for a given noise level $\sigma$, a statistically averaged measurement $\langle...\rangle_{\sigma}$ over frequency fluctuations can be formally written:

\begin{equation}
\begin{split}
\label{AltSusceptibility}
\chi_{\sigma}(\omega)&=\langle\chi_0(\omega-\delta\Omega)\rangle_{\sigma}\\&=\int_{-\infty}^{+\infty}\chi_0(\omega-\delta\Omega)\rho_{\sigma}(\delta\Omega)\mathrm{d}\delta\Omega,
\end{split}
\end{equation}
hence the convolution picture for the susceptibility. 
Eq. (\ref{AltSusceptibility}) only assumes that fluctuations are slow enough to validate this adiabatic picture, and fast enough to be fully integrated in the measurement.
The formalism can be applied to the Telegraph noise of Ref. \cite{Telegraph}, when the switching rate is slow, with a distribution $\rho_\sigma$ consisting of two Delta functions separated by the switching frequency span.

In the convolution picture, it appears that the frequency fluctuations do not constitute an additional dissipative channel. Indeed, the distribution $\rho_{\sigma}$ is normalized to 1 independently of the noise level $\sigma$. Then, using Eq. (\ref{AltSusceptibility}):

\begin{align*}
&\int_{-\infty}^{+\infty}\mathrm{Im}\left[\chi_{\sigma}(\omega)\right]\mathrm{d}\omega\\
&=\int_{-\infty}^{+\infty}\mathrm{Im}\left[\chi_{0}(u)\right]\mathrm{d}u\int_{-\infty}^{+\infty}\rho_{\sigma}(\delta\Omega)\mathrm{d}\delta\Omega\\
&=\int_{-\infty}^{+\infty}\mathrm{Im}\left[\chi_{0}(\omega)\right]\mathrm{d}\omega.
\end{align*}

Likewise:

\begin{align*}
\int_{-\infty}^{+\infty}\mathrm{Re}\left[\chi_{\sigma}(\omega)\right]\mathrm{d}\omega
&=\int_{-\infty}^{+\infty}\mathrm{Re}\left[\chi_{0}(\omega)\right]\mathrm{d}\omega=0.
\end{align*}
In other words, the area under X and Y quadratures curves, i.e. the energy of the detected signal is preserved, independently of the frequency fluctuations.

\section{FREQUENCY FLUCTUATIONS SPECTRAL DENSITY}
\label{AppendixSpectrum}

Our formalism does not require to know the frequency noise spectral density: only its integrated value is needed. Yet we can infer this spectrum, since our (Gaussian) noise source and our filter are calibrated.

All parts of the filtering system are obviously linear and causal. Then, while theoretical calculations are made over both negative and positive frequencies, experimental values and curves are given only for calculations over positive frequencies. The causality assumption makes the latter differ only by a factor of 2 from theory.

The random variables $\delta V_g(t)$ and $\delta V_g(t+\tau)$ are both Gaussian, centered, with the same standard deviation value $\sigma_V$. Using Wick formula (a brief demonstration is given in the supplementary material), one derives the autocorrelation function for the frequency fluctuations, as a function of the autocorrelation for gate voltage fluctuations:

\begin{align*}
\langle\delta\Omega(t)\delta\Omega(t+\tau)\rangle &=\alpha^2\langle\delta V_g(t)\delta V_g(t)\delta V_g(t+\tau)\delta V_g(t+\tau)\rangle\\
&=\alpha^2\sigma_V^4+2\alpha^2\langle\delta V_g(t)\delta V_g(t+\tau)\rangle^2\\
&=\frac{\sigma^2}{2}+2\alpha^2\langle\delta V_g(t)\delta V_g(t+\tau)\rangle^2.
\end{align*}

Remarkably, the frequency fluctuations autocorrelation function does not go to zero at long times. This is the result of the non-zero mean value of the frequency noise, which is as a matter of fact the constant value in the final expression of the autocorrelation. Using Eq. (\ref{Wiener-Khinchin}) one writes the frequency fluctuations spectrum as a function of the gate voltage fluctuations spectrum $S_V$:

\begin{equation}
\label{Spectrum}
S_{\Omega}(\omega)=\frac{\sigma^2}{2}\delta(\omega)+\frac{\alpha^2}{\pi}\int_{-\infty}^{+\infty}S_V(\omega')S_V(\omega-\omega')\mathrm{d}\omega',
\end{equation}

One can recognize in the last term the auto-convolution of the voltage fluctuations spectrum and then reconstruct the frequency fluctuations spectrum: the gate electrode is fed with a white noise standard deviation value $\sigma_A$ over an equivalent bandwith (over positive frequencies) $\mathrm{BW}_A\simeq 2\pi\times 9~\mega\hertz$. For the reasons explained in the main article, this noise is filtered at the output of the AWG. The power gain curve of the filter is fit with an algebraic function $F(\omega)$ given in Supplemental Material \cite{SI}. The integral calculated from the algebraic fit (over positive frequencies) yields the equivalent bandwith of this filter $\mathrm{BW}_F\simeq 2\pi\times 1.2~\hertz.$

\begin{figure}[h!]
\includegraphics[scale=0.8]{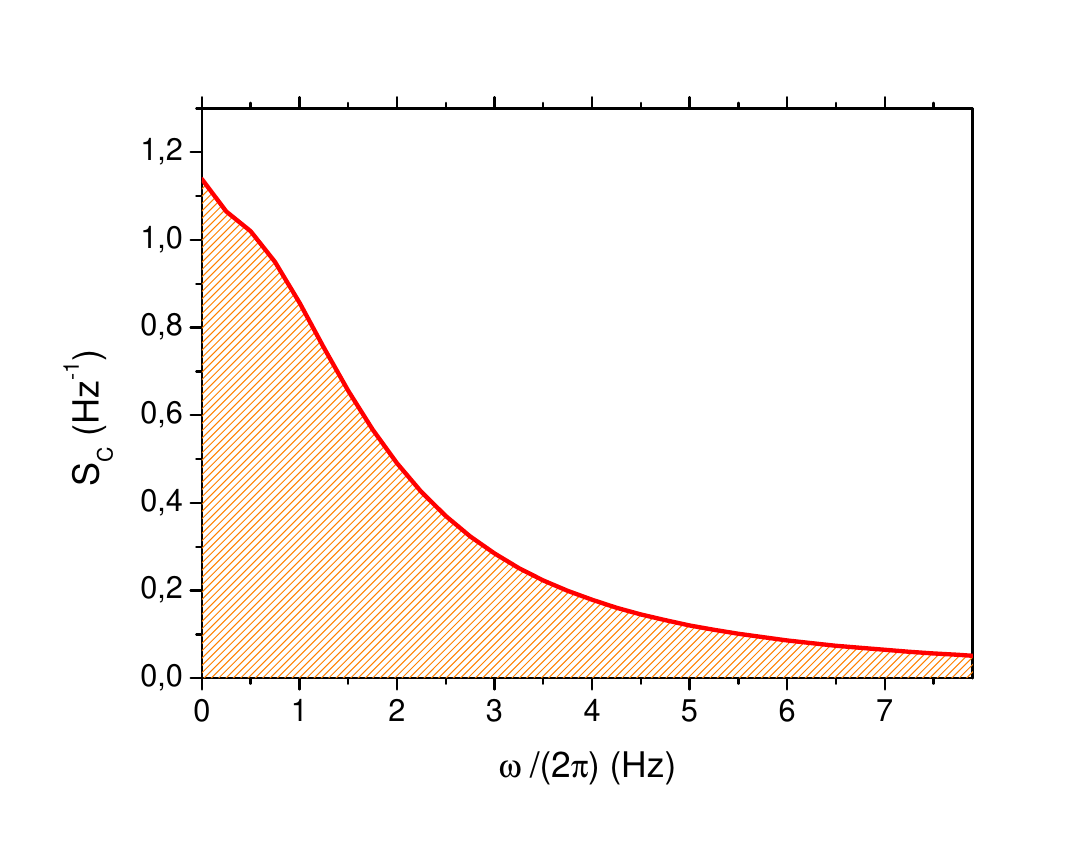}
\caption{(Color online) Convolution term $S_C(\omega)$ of the frequency fluctuations spectrum, with the area under the curve filled in orange.}
\label{fig:FluctSpectrum}
\end{figure}

The filtered noise is then amplified over its whole frequency range with a power gain $G=2.5\times 10^{7}$. Taking into account again causality and linearity of the injection chain, one can recalculate the standard deviation value of the gate voltage noise $\sigma_V$:

\begin{equation}
\label{sigmaVgate}
\sigma_V^2=\frac{\mathrm{BW}_F}{\mathrm{BW}_A}G\sigma_A^2.
\end{equation}

For an injected noise of standard deviation value $\sigma_A=0.25~\volt$ at the AWG output port, one gets $\sigma_V=460~\milli\volt$, which matches within $10\%$ the measured value of $490~\milli\volt$. The spectrum at the filter output, including the amplification factor $G$ is then:

\begin{equation}
\label{VoltSpectrum}
S_V(\omega)=\frac{\pi\sigma_A^2G}{\mathrm{BW}_A}F(\omega)=\frac{\pi\sigma_V^2}{\mathrm{BW}_F}F(\omega).
\end{equation}

Then the frequency fluctuations spectrum reads:

\begin{multline}
\label{SpectrumDetails}
S_{\Omega}(\omega)=\frac{\sigma^2}{2}\delta(\omega)\\+\frac{\pi\alpha^2\sigma_V^4}{\mathrm{BW}_F^2}\underbrace{\int_{-\infty}^{+\infty}F(\omega')F(\omega-\omega')\mathrm{d}\omega'}_{S_C(\omega)}.
\end{multline}

From experimental parameters we can thus quantitatively infer the frequency fluctuations spectrum. The convolution part $S_C(\omega)$ of this spectrum is given in Fig. \ref{fig:FluctSpectrum}.

The convolution weighs the spectrum at low frequencies, with maximum at zero. Through this definition we can verify that the integrated value of the spectrum gives again $3\sigma^2/2$:

\begin{align*}
&\frac{1}{2\pi}\int_{-\infty}^{+\infty}S_{\Omega}(\omega)\mathrm{d}\omega \\
&=\frac{\sigma^2}{2}+\frac{\alpha^2\sigma_V^4}{2\mathrm{BW}_F^2}\times 4\mathrm{BW}_F^2\\
&=\frac{3\sigma^2}{2}.
\end{align*}

\section{SHAPE ANALYSIS}
\label{AppendixShape}

\begin{figure}
\includegraphics[scale=0.8]{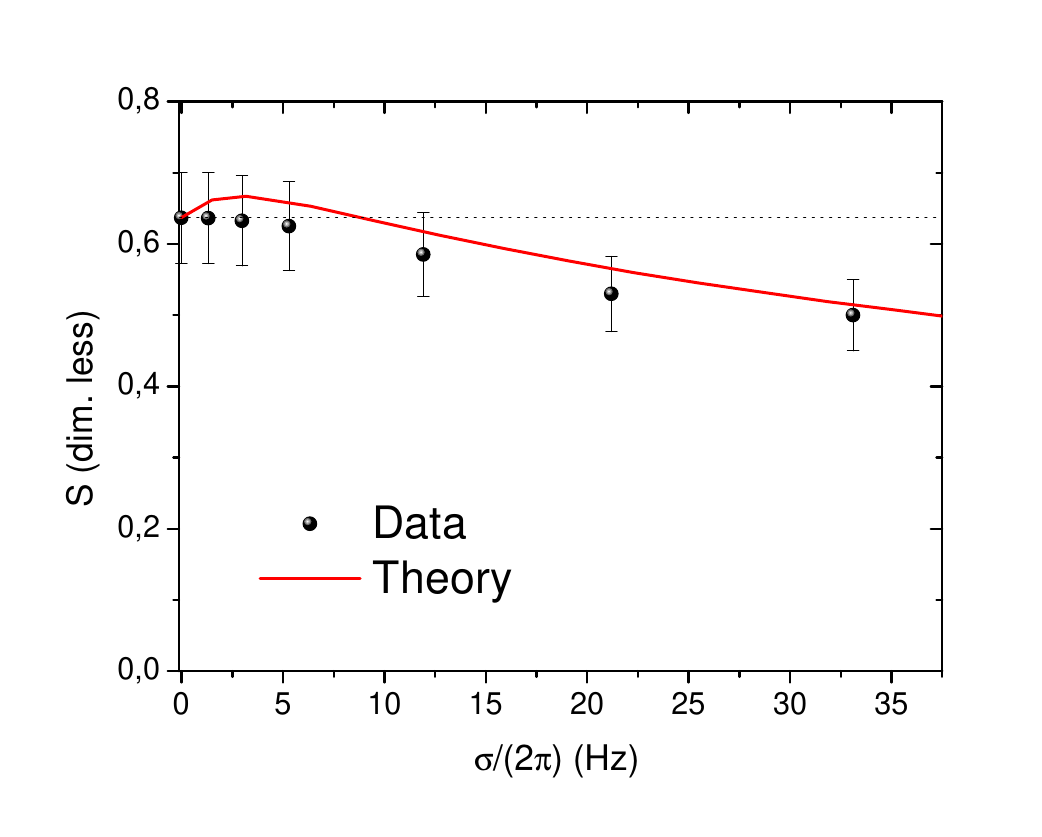}
\caption{(Color online) Shape factor as a function of the level of noise applied. Black dots are obtained from measurements, while the red curve is the application of Eq. (\ref{shape}). The dotted line is the Lorentzian case.}
\label{fig:Shape}
\end{figure}

The resonance lines obtained under frequency fluctuations are clearly not Lorentzian. Yet we can approximate them as such to quantify decoherence in our system, and thus extract their full width at half maximum (FWHM), for not too large noise levels. Let us define a shape factor $S$ as:

\begin{equation}
\label{shape}
S(\sigma)=\frac{2\pi W(\sigma)x_{max}(\sigma)}{\int\mathrm{Im}\left[\chi_0(\omega)\right]F_0\mathrm{d}\omega},
\end{equation}

\begin{figure}
			 \includegraphics[scale=0.7]{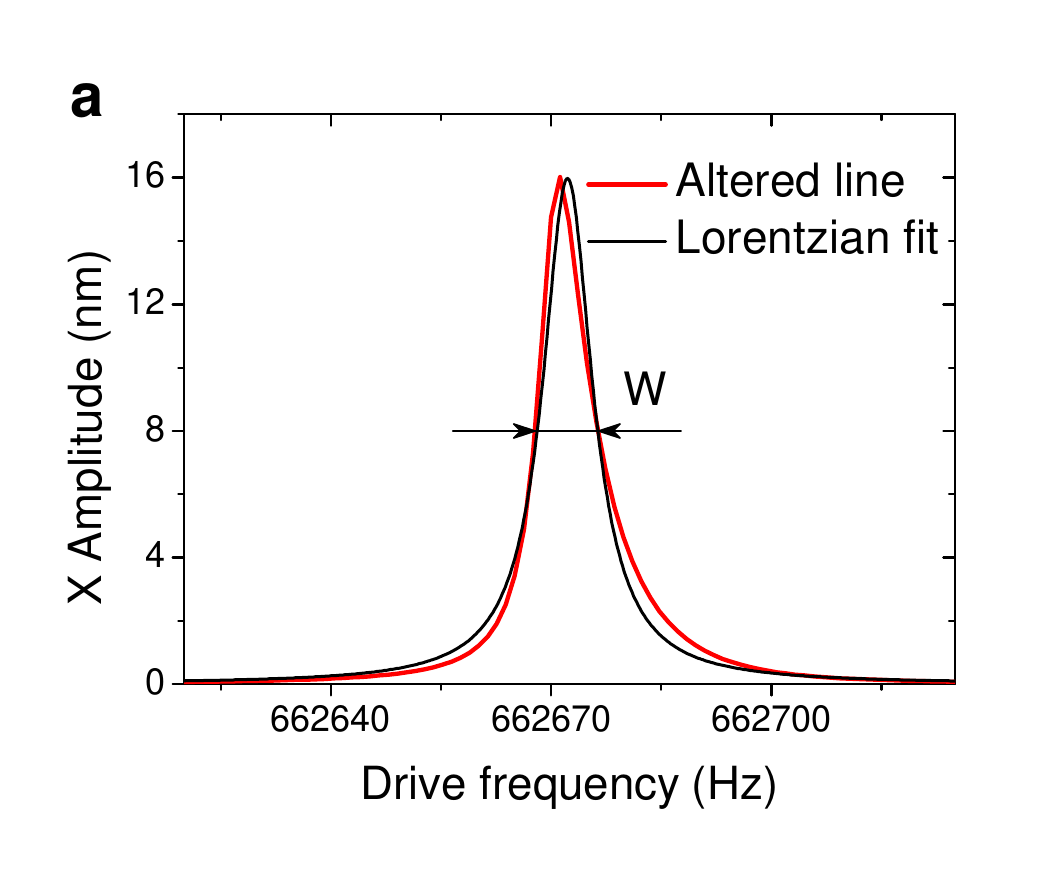}	
			  \includegraphics[scale=0.7]{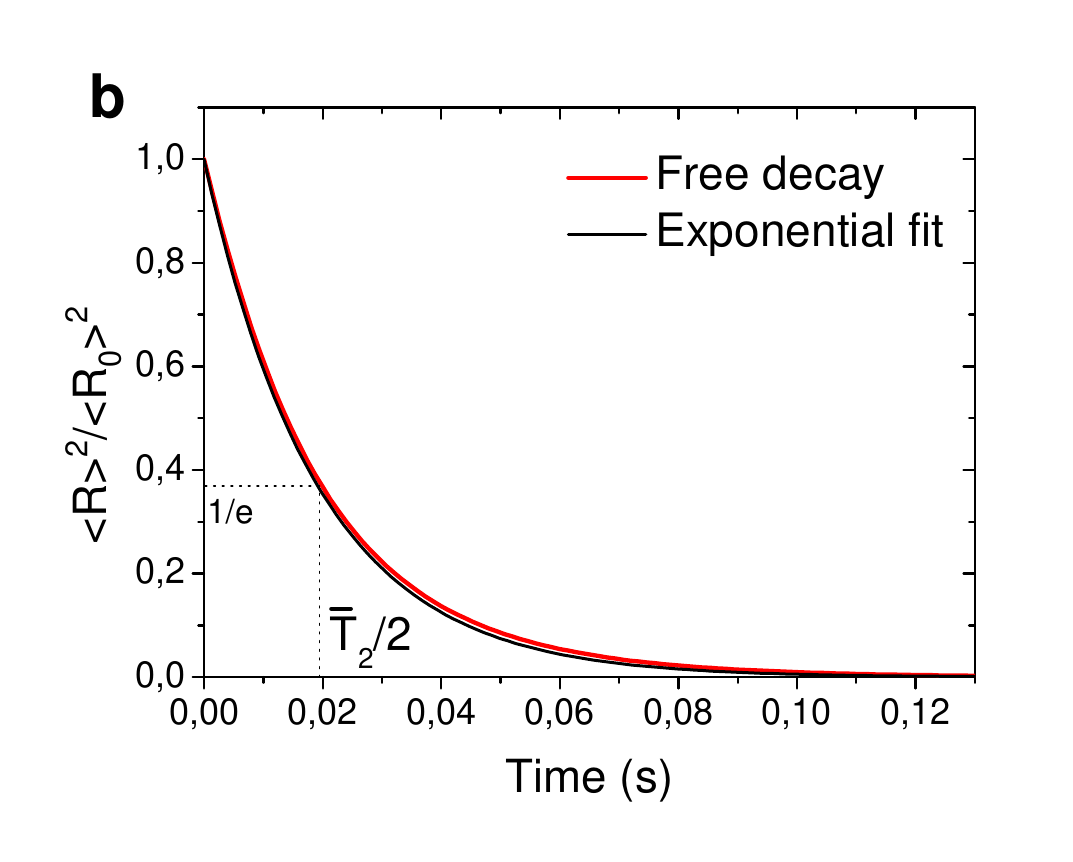} 
\caption{(Color online) (a) Frequency-domain response in the X quadrature at $\sigma=2\pi\times 5.1~\hertz$ (red curve), with a Lorentzian fit of FWHM $W=8.3~\hertz$ (black curve). (b) Time-domain response of $\langle R\rangle^2$ normalized to the steady state resonant amplitude (red curve), with an exponential fit of time constant $\overline{T}_2/2 = 0.02~$s (black curve).}
\label{fig:ShapeApprox}
\end{figure}

with $W$ in $\hertz$ units, hence the $2\pi$ factor. Here we take advantage of the fact that the area of a true Lorentzian of FWHM $W=W_0$ and height $x_{max}=x_0$ is $\pi^2 W_0x_0$. Then, for no noise applied, $S$ is simply $2/\pi$. Fig. \ref{fig:Shape} displays the evolution of the shape factor as a function of $\sigma$, with experimental data matching Eq. (\ref{shape}) within $\pm~10\%$. In our experimental range, $S$ deviates at most by $20\%$ from its noise-free value which corresponds to the Lorentzian shape. This implies that a simple Lorentzian analysis in frequency domain, and a single exponential fit in time domain, are rather accurate. To illustrate this we plot in Fig. \ref{fig:ShapeApprox} the actual calculated frequency and time domain curves (red) for $\sigma=2\pi\times 5.1~\hertz$ together with the simple fits (black). 
The time domain curve is easily obtained by writing:
\begin{equation}
\label{time-domain}
\langle\mathrm{R}(t)\rangle_N^2= \langle \mathrm{X}(t)\rangle_N^2+\langle \mathrm{Y}(t)\rangle_N^2 ,
\end{equation}
with
\begin{align*}
&\mathrm{X}(t)=\exp (-t/T_1) \big[\mathrm{X}(\omega_0-\delta\Omega) \cos \big((\Delta+\delta\Omega) t\big)\\&-\mathrm{Y}(\omega_0-\delta\Omega)\sin\big((\Delta+\delta\Omega) t\big)\big],\\
&\mathrm{Y}(t)=\exp (-t/T_1) \big[\mathrm{Y}(\omega_0-\delta\Omega) \cos \big((\Delta+\delta\Omega) t\big)\\&+\mathrm{X}(\omega_0-\delta\Omega)\sin\big((\Delta+\delta\Omega) t\big)\big],
\end{align*}
where $\Delta$ is the frequency of the beating between the mechanics and the local oscillator, and $\langle ... \rangle_N \approx \langle ... \rangle_{\sigma} $ by construction.
For this particular case, $\sigma/(2\pi) $ is of the order of the fluctuation-free linewidth $W_0$.
As can be seen on the graphs, while the shapes are different the key parameters $W$ and $\overline{T}_2$ are essentially the same for the two types of curves.
We furthermore verify $\overline{T}_2/2 \approx (2\pi W)^{-1}$ when the applied noise is not too large.
An equivalent alternative way of studying the shape of the resonance is to define the integrated width ${\cal I} = \int\mathrm{Im}\left[\chi_0(\omega)\right]F_0\mathrm{d}\omega / x_{max}$, following the procedure of Ref. \cite{NatNanoMoser}. Indeed, the shape factor is nothing but the ratio of the FWHM to the integrated width: while the resonance peak remains close to a Lorentzian, ${\cal I} \approx \frac{\pi}{2} \, 2\pi W$.

\vspace*{5 mm}
Corresponding Author: E. C., $\underline{\mbox{eddy.collin@neel.cnrs.fr}}$

\bibliographystyle{ieeetran}

\end{document}